\documentclass[preprint,aps]{revtex4-1}
\usepackage{graphicx}
\usepackage{hyperref}
\usepackage{epstopdf}
\usepackage{slashed}
\usepackage{rotating}
\usepackage{float}
\newcommand{\be}{\begin{equation}}
\newcommand{\ee}{\end{equation}}
\newcommand{\bea}{\begin{eqnarray}}
\newcommand{\eea}{\end{eqnarray}}
\newcommand{\ba}{\begin{array}}
\newcommand{\ea}{\end{array}}

\begin{document}
\title{Transition magnetic moments of $J^P=\frac{3}{2}^+$ decuplet to $J^P=\frac{1}{2}^+$ octet baryons in the chiral constituent quark model}

\author{Harleen Dahiya}
\affiliation{Department of Physics, Dr. B.R. Ambedkar National
Institute of Technology, Jalandhar, 144011, India}

\begin{abstract}

In light of the developments of the chiral constituent quark model ($\chi$CQM) in studying low energy hadronic matrix elements of the ground-state baryons, we have extended this model to investigate their transition properties. The magnetic moments of transitions from the $J^P=\frac{3}{2}^+$ decuplet to $J^P=\frac{1}{2}^+$ octet baryons have been calculated with explicit valence quark spin, sea quark spin  and sea quark orbital angular momentum contributions. Since the experimental data is available for only a few transitions, we have compared our results with the results of other available models. The implications of other complicated effects such as chiral symmetry breaking and SU(3) symmetry breaking arising due to confinement of quarks have also been discussed.

\end{abstract}
\maketitle
\section{Introduction}
\label{intro}

Understanding the internal structure of hadrons within the nonperturbative regime of quantum chromodynamics (QCD) is one of the most challenging area in  theory as well as experiment.   The electromagnetic properties,  obtained from the measurements of electromagnetic Dirac and Pauli form factors, are further related to the static low-energy observables like masses, charge radii, magnetic moments, etc.. It constitutes one of the most promising area and can provide valuable insight into the underlying dynamics and the nonperturbative aspects of QCD. At present, electromagnetic form factors have been precisely obtained for the case of nucleons \cite{emc,smc,adams,e866,hermes,pdg} whereas, for other baryons, the experimental data are available only for magnetic moments.

Magnetic moment of baryons is one of the most important quantity in scrutinizing the structure and the properties of light baryons. Continuous theoretical efforts are being
made to investigate the magnetic moments and the calculations have benefited a lot from the information being made available through the experiments. At present, the magnetic moments of the $J^P=\frac{1}{2}^+$ octet baryons (except $\Sigma^0$) have been accurately measured experimentally \cite{pdg}. Our information about the $J^P=\frac{3}{2}^+$ decuplet baryons is however limited to only $\Delta^+$ and $\Omega^-$ because of the difficulty in measuring their properties experimentally on account of their short lifetimes.

Further, the low-lying baryon decuplet to octet electromagnetic transitions play a very important role in probing the internal spin structure as well as the deformation of the octet and decuplet baryons. The $\Delta(1232)$ resonance is the lowest-lying excited state of the nucleon in which the search for transition amplitudes from the spin-parity selection rules has been carried out.  The $\Delta^+ \rightarrow p \gamma$  transition amplitude contains the magnetic dipole moment ($G_{M1})$, the electric quadrupole moment ($G_{E2}$), and the Coulumb quadrupole moment ($G_{C2})$. The information on magnetic moment is obtained from $G_{M1}$ amplitude, whereas $G_{E2}$ and $G_{C2}$ amplitudes give us information on the intrinsic quadrupole moment. In spite of considerable efforts put in over the past few decades to determine the magnetic moments of the octet as well as the decuplet baryons, the decuplet to octet transition magnetic moments are less well-known. The transition magnetic moments are difficult to understand since the decuplet baryons have very short lifetime and also the magnetic moments receive contributions from various interrelated effects, for example, spin and orbital angular momentum contributions, relativistic and exchange current effects,
spin-0 meson cloud contributions, effect of the confinement on quark masses, etc..

The magnetic moments of $J^P=\frac{1}{2}^+$ octet and $J^P=\frac{3}{2}^+$ decuplet baryons have been extensively calculated theoretically using numerous different approaches. The approaches include a SU(6) symmetric naive quark model (NQM) \cite{nqm,mgupta}, nonrelativistic quark  model \cite{nrqm} relativistic quark model \cite{rqm},  QCD-based quark model \cite{pha,slaughter}, chiral perturbation theory \cite{chipt} QCD string approach
\cite{kerbikov},  light cone QCD sum rule \cite{aliev}, QCD sum rule 
\cite{zhu},  hypercentral model \cite{pcvinod},  Skyrme model \cite{schwesinger}, soliton model
\cite{ledwig,kim},  large-$N_c$ chiral perturbation theory
\cite{mendieta}, lattice QCD
\cite{boinepalli,lattice2}, chiral quark model with exchange currents \cite{buchmann}.
These studies indicate the growing interest in this field. The study on the transition magnetic moments is rather limited and a few attempts have been made in chiral perturbation theory ($\chi$PT) \cite{chipt-transition}, light cone QCD sum rules  and light cone QCD (LCQCDSR) \cite{aliev-transition}, large-$N_c$ chiral perturbation theory (Large $N_C$ PT) \cite{1largeNc-transition,2largeNc-transition}, relativistic quark model (Rel-QM)\cite{rel-transition},  cloudy bag model (CBM) \cite{cloudy-transition},  Skyrme model (SM) \cite{skyrme-transition} , QCD sum rules (QCDSR) \cite{1qcdsr-transition, 2qcdsr-transition},  lattice
QCD \cite{lattice-transition}, chiral quark model ($\chi$QM) \cite{chiqm-transition} effective mass quark model (EMQM) \cite{emqm-transition}, meson cloud model (MCM) \cite{meson-cloud-transition} , U-spin \cite{uspin-transition} etc..

One of the important model which finds application in the nonperturbative regime of QCD is the chiral constituent quark model ($\chi$CQM) \cite{manohar,eichten,cheng} where chiral symmetry breaking and its spontaneous breaking is implemented.  The $\chi$CQM uses the effective interaction Lagrangian approach of the strong interactions,  where, the important phenomenon of quark-antiquark excitations is included. This results in the presence of the meson cloud at low energies where the effective degrees of freedom are the valence quarks and the internal Goldstone bosons (GBs),
which are coupled to the valence quarks \cite{cheng,johan,song,hd,hds}. This perspective is in common with the modern effective field theory approaches. The $\chi$CQM  has successfully been applied to calculate the  spin and flavor distribution functions including the strangeness content of the nucleon \cite{song,hd},
weak vector and axial-vector form factors \cite{nsweak}, nucleon structure functions and longitudinal spin asymmetries \cite{hd-spinasymm-2016}, electromagnetic and axial-vector form factors of the quarks and nucleon \cite{hd-em-2017}, charge radii and quadrupole moment \cite{charge-radii}. The magnetic moments of octet baryons, the transition within the octet baryons $\Sigma \rightarrow \Lambda$ and the Coleman-Glashow sum rule have already been calculated \cite{hdmagnetic}. The work was further extended to the calculations of the magnetic moments of decuplet baryons  \cite{hd-decuplet}, the magnetic moments baryon resonances \cite{nres}, magnetic moments of $\Lambda$ resonances \cite{torres} etc..

Considering the above developments of the $\chi$CQM in studying
low energy hadronic matrix elements of the ground-state baryons, it becomes desirable to extend this model to investigate their transition properties. We will calculate the magnetic moments of transitions from the $J^P=\frac{3}{2}^+$ decuplet to $J^P=\frac{1}{2}^+$ octet baryons. Taking benefit from the earlier
studies of $J^P=\frac{3}{2}^+$ decuplet  and $J^P=\frac{1}{2}^+$ octet baryons \cite{hdmagnetic}, the explicit contributions coming from the valence quarks, quark sea
polarization, and its orbital angular momentum have been calculated.
The implications of other complicated effects such as chiral symmetry breaking and SU(3) symmetry breaking  arising due to confinement of quarks have also been discussed.

\section{Transition Magnetic moments} \label{magmom}

In this section, we calculate the transition magnetic moments for
the radiative decays $B_i \rightarrow B_f + \gamma,$ where $B_i$ and
$B_f$ are the initial and final baryons. Since the $M1$ transition involves the quark magnetic moments, it can lead to the transition between the spin
${\frac{3}{2}}^+$ decuplet to the spin ${\frac{1}{2}}^+$ octet. We consider here the magnetic moments of the spin ${\frac{3}{2}}^+ \to {\frac{1}{2}}^+$ transitions. In the present calculations we have considered only the $S_z=\frac{1}{2}$ spin projection for the ${\frac{3}{2}}^+$ decuplet  as the matrix elements for other spin projections will come out to be zero.

\vspace {0.3cm}
The transition magnetic moment can be calculated
from the matrix element \be \mu \left (B_{\frac{3}{2}^+} \rightarrow B_{\frac{1}{2}^+}\right )=\left\langle  B_{\frac{1}{2}^+} , S_z=\frac{1}{2} \right |\mu_z \left| B_{\frac{3}{2}^+} , S_z=\frac{1}{2}  \right\rangle  \,,\ee 
where $\mu_z$ corresponds to the magnetic moment operator, $\left | B_{\frac{1}{2}^+} \right\rangle $ and $\left | B_{\frac{3}{2}^+} \right\rangle $ correspond to the spin-flavor wavefunctions of the octet and decuplet baryons respectively expressed as 
\bea \left | B_{\frac{3}{2}^+} \right\rangle 
&\equiv& \left|10, {\frac{3}{2}}^+ \right \rangle = \chi^{s} \phi^s \,,\label{decuplet} \\
\left | B_{\frac{1}{2}^+} \right\rangle  &\equiv&
\left|8, {\frac{1}{2}}^+ \right \rangle = \frac{1}{\sqrt 2}(\chi^{'} \phi^{'} +
\chi^{''} \phi^{''})\,. \label{octet}  \eea
The spin wavefunctions ($\chi^{s}$ for the case of decuplet baryons and $\chi^{'}$ and $\chi^{''}$ for the case of octet baryons)   are expressed as
\bea \chi^{s} &=&  \uparrow \uparrow \uparrow\,, \nonumber \\
\chi^{'} &= & \frac{1}{\sqrt 2}(\uparrow \downarrow \uparrow
-\downarrow \uparrow \uparrow)\,, \nonumber \\ \chi^{''} &=&
\frac{1}{\sqrt 6} (2\uparrow \uparrow \downarrow -\uparrow
\downarrow \uparrow -\downarrow \uparrow \uparrow)\,. \eea

The flavor wavefunctions
$\phi^{s}$ for the decuplet baryons of the types $ B_{\frac{3}{2}^+}(Q_1Q_1Q_1)$,
$ B_{\frac{3}{2}^+}(Q_1Q_1Q_2)$ and $ B_{\frac{3}{2}^+}(Q_1Q_2Q_3)$ are  respectively expressed as \bea
\phi^{s}_{B^*} &=& Q_1Q_1Q_1\,, \nonumber \\
\phi^{s}_{B^*} &=& \frac{1}{\sqrt 3}(Q_1Q_1Q_2 + Q_1Q_2Q_1 + Q_2Q_1Q_1)\,, \nonumber \\
\phi^{s}_{B^*} &=& \frac{1}{\sqrt 6}(Q_1Q_2Q_3 + Q_1Q_3Q_2 + Q_2Q_1Q_3 +Q_2Q_3Q_1 + Q_3Q_1Q_2 + Q_3Q_2Q_1)\,, \eea
whereas the   flavor wavefunctions
$\phi^{'}$ and $\phi^{''}$ for the octet baryons of the type
$ B_{\frac{1}{2}^+}(Q_1Q_1Q_2)$ are \bea \phi^{'}_B &=& \frac{1}{\sqrt 2}(Q_1Q_2Q_1-Q_2Q_1Q_1)\,,
\nonumber \\ \phi^{''}_B &=& \frac{1}{\sqrt 6}(2Q_1Q_1Q_2- Q_1Q_2Q_1-Q_2Q_1Q_1)\,, \eea
where $Q_1$, $Q_2$, and $Q_3$ correspond to any of the $u$, $d$, and $s$
quarks. For the case of $\Lambda(uds)$ and $\Sigma^0(uds)$, the
wavefunctions are given as \bea \phi^{'}_{\Lambda} &=&
\frac{1}{2\sqrt 3}(usd+sdu-sud-dsu-2uds-2dus)\,, \nonumber \\
\phi^{''}_{\Lambda} &=& \frac{1}{2}(sud+usd-sdu-dsu)\,, \nonumber \\
\phi^{'}_{\Sigma^0}&=& \frac{1}{2}(sud+sdu-usd-dsu)\,, \nonumber \\
\phi^{''}_{\Sigma^0} &=&\frac{1}{2 \sqrt
	3}(sdu+sud+usd+dsu-2uds-2dus)\,. \eea 
The details of the spatial wave functions ($\psi^{s}, \psi^{'}, \psi^{''})$  can be found in Ref. \cite{yaouanc}.

The magnetic moment of a given baryon in the $\chi$CQM receives
contribution from the valence quark spin, sea quark spin  and sea quark orbital angular momentum. The total magnetic moment is expressed as \be  \mu \left (B_{\frac{3}{2}^+} \rightarrow B_{\frac{1}{2}^+}\right )_{{\rm Total}}= \mu\left (B_{\frac{3}{2}^+} \rightarrow B_{\frac{1}{2}^+}\right )_{{\rm V}}+\mu\left (B_{\frac{3}{2}^+} \rightarrow B_{\frac{1}{2}^+}\right )_{{\rm S}} + \mu\left (B_{\frac{3}{2}^+} \rightarrow B_{\frac{1}{2}^+}\right )_{{\rm O}}\,, \label{totalmag} \ee
where $\mu\left (B_{\frac{3}{2}^+} \rightarrow B_{\frac{1}{2}^+}\right )_{{\rm V}}$ and $\mu\left (B_{\frac{3}{2}^+} \rightarrow B_{\frac{1}{2}^+}\right )_{{\rm S}}$ are the magnetic moment contributions of the valence quarks and the sea quarks respectively coming from their spin polarizations, whereas  $\mu\left (B_{\frac{3}{2}^+} \rightarrow B_{\frac{1}{2}^+}\right )_{{\rm O}}$ is the magnetic moment contribution due to the rotational
motion of the two bodies constituting the sea quarks and Goldstone boson (GB)  and referred to as the orbital
angular momentum contribution of the quark sea \cite{cheng}.

In terms of quark magnetic moments and spin polarizations, the
valence spin $\mu\left (B_{\frac{3}{2}^+} \rightarrow B_{\frac{1}{2}^+}\right )_{{\rm V}}$, sea spin $\mu\left (B_{\frac{3}{2}^+} \rightarrow B_{\frac{1}{2}^+}\right )_{{\rm S}}$, and sea orbital $\mu\left (B_{\frac{3}{2}^+} \rightarrow B_{\frac{1}{2}^+}\right )_{{\rm O}}$ contributions can be defined as
 \bea
\mu\left (B_{\frac{3}{2}^+} \rightarrow B_{\frac{1}{2}^+}\right )_{{\rm V}} &=& \sum_{q=u,d,s}{\Delta q\left ({\frac{3}{2}^+} \rightarrow {\frac{1}{2}^+}\right )_{{\rm V}}\mu_q}\,,\label{mag-val}\\
\mu\left (B_{\frac{3}{2}^+} \rightarrow B_{\frac{1}{2}^+}\right )_{{\rm S}} &=& \sum_{q=u,d,s}{\Delta q\left ({\frac{3}{2}^+} \rightarrow {\frac{1}{2}^+}\right )_{{\rm S}}\mu_q}\,,\label{mag-sea} \\ \mu\left (B_{\frac{3}{2}^+} \rightarrow B_{\frac{1}{2}^+}\right )_{{\rm O}} &=& \sum_{q=u,d,s}{\Delta q\left ({\frac{3}{2}^+} \rightarrow {\frac{1}{2}^+}\right )_{{\rm V}}~\mu(q_{+} \rightarrow )} \,,\label{mag-orbit}
 \eea
 where
$\mu_q= \frac{e_q}{2 M_q}$ ($q=u,d,s$) is the quark magnetic moment in the units of $\mu_N$ (nuclear magneton), $\Delta q\left ({\frac{3}{2}^+} \rightarrow {\frac{1}{2}^+}\right )_{{\rm V}}$
and $\Delta q\left ({\frac{3}{2}^+} \rightarrow {\frac{1}{2}^+}\right )_{{\rm S}}$ are the valence and sea quark spin polarizations respectively,
$\mu(q_{+} \rightarrow )$ is the orbital moment for any chiral
fluctuation, $e_q$ and $M_q$ are the electric charge and the mass,
respectively, for the quark $q$.

The spin structure of a decuplet to octet transition matrix element is defined as
 \be \left\langle  B_{\frac{1}{2}^+} , S_z=\frac{1}{2} \right |{\cal N}\left (B_{\frac{3}{2}^+} \rightarrow B_{\frac{1}{2}^+}\right )\left| B_{\frac{3}{2}^+} , S_z=\frac{1}{2}  \right\rangle  \,,\ee 
here the number operator ${\cal N}\left (B_{\frac{3}{2}^+} \rightarrow B_{\frac{1}{2}^+}\right )$  measures the number of quarks with spin up ($\uparrow$) or down ($\downarrow$) in the transition ${\frac{3}{2}^+} \rightarrow {\frac{1}{2}^+}$, for example,
\be
{\cal N}\left (B_{\frac{3}{2}^+} \rightarrow B_{\frac{1}{2}^+}\right )
=\sum_{q=u,d,s}\left(N_{q^{\uparrow}\left (B_{\frac{3}{2}^+} \rightarrow B_{\frac{1}{2}^+}\right )}+ N_{q^{\downarrow}\left (B_{\frac{3}{2}^+} \rightarrow B_{\frac{1}{2}^+}\right )}\right )\,. \label{number}
\ee
This can be used to calculate the quark spin polarizations (for $q=u, d, s$) for a given transition 
\be \Delta q\left (B_{\frac{3}{2}^+} \rightarrow B_{\frac{1}{2}^+}\right )= N_{q^{\uparrow}\left (B_{\frac{3}{2}^+} \rightarrow B_{\frac{1}{2}^+}\right )}- N_{q^{\downarrow}\left (B_{\frac{3}{2}^+} \rightarrow B_{\frac{1}{2}^+}\right )} \,.\ee

The valence quarks spin polarizations $\Delta q\left ({\frac{3}{2}^+} \rightarrow {\frac{1}{2}^+}\right )_{{\rm V}}$ for a given baryon transition can be
calculated using the SU(6) spin-flavor wave functions defined in
Eqs. (\ref{decuplet}) and (\ref{octet}). Using these, the magnetic moment contribution coming from the valence quarks can  be calculated from Eq. (\ref{mag-val})  and have been summarized in Table \ref{val-trans} for all the decuplet to octet transitions.

\begin{table}
	\begin{center}
		\tabcolsep 2.5 mm {\renewcommand{\arraystretch}{1.5}
			\begin{tabular}{|c|c|}  \hline
				$\frac{3}{2}^+ \rightarrow \frac{1}{2}^+$ transition &  Valence transition magnetic moments \\ \hline 
				
				$\mu (\Delta \rightarrow p )_{{\rm V}}$  & $\frac{2\sqrt{2}}{3}\mu_{u} -
				\frac{2\sqrt{2}}{3}\mu_{d}$\\
				
			$\mu (\Sigma^{*+} \rightarrow \Sigma^{+} )_{{\rm V}}$  & $\frac{2\sqrt{2}}{3}\mu_{u}-
				\frac{2\sqrt{2}}{3}\mu_{s}$\\
				
					$\mu ({\Sigma^{*0}\rightarrow\Sigma^{0}})_{{\rm V}}$  & $\frac{\sqrt{2}}{3}
				\mu_{u}+ \frac{\sqrt{2}}{3}\mu_{d}- \frac{2\sqrt{2}}{3}\mu_{s}$ \\
				
					$\mu({\Sigma^{*0} \rightarrow \Lambda})_{{\rm V}}$  &$\sqrt{\frac{2}{3}}\mu_{u}-
				\sqrt{\frac{2}{3}}\mu_{d}$\\

			$\mu({\Sigma^{*-}\rightarrow\Sigma^{-}})_{{\rm V}}$ & $\frac{2\sqrt{2}}{3}\mu_{d} -
				\frac{2\sqrt{2}}{3}\mu_{s}$\\
				
				$\mu({\Xi^{*0}\rightarrow\Xi^{0}})_{{\rm V}}$  & $\frac{2\sqrt{2}}{3}\mu_{u}-
				\frac{2\sqrt{2}}{3}\mu_{s}$\\
				
			$\mu({\Xi^{*-}\rightarrow\Xi^{-}})_{{\rm V}}$  &$\frac{2\sqrt{2}}{3}\mu_{d}-
				\frac{2\sqrt{2}}{3}\mu_{s}$\\
				
			 \hline
\end{tabular}
\caption{Valence transition magnetic moments for ${\frac{3}{2}}^+ \to {\frac{1}{2}}^+$ transitions.}\label{val-trans} }
\end{center}
\end{table}

For the calculation of the  sea quarks spin polarizations $\Delta q\left ({\frac{3}{2}^+} \rightarrow {\frac{1}{2}^+}\right )_{{\rm S}}$ for a given baryon transition,  we will use the basic idea of the chiral constituent quark model ($\chi$CQM)  \cite{manohar} where the set of internal
GBs couple directly to the valence quarks in the
interior of hadron and we have 
\be
q^{\uparrow \downarrow} \to  P(q,GB) q^{\uparrow \downarrow} + P(q^{\uparrow \downarrow},GB). \label{sea-q}
\ee
Here the transition probability of the emission of a GB from
any of the $q$ quark $ P(q,GB)$  and the transition
probability of the $q^{\uparrow \downarrow}$ quark $P(q^{\uparrow \downarrow},GB)$ can be calculated from the effective
Lagrangian describing interaction between quarks and GBs expressed as
\be {\cal
	L}_{{\rm int}} = c_8 {\bar \psi}\left( \Phi + P_{\eta'}
\frac{\eta'}{\sqrt 3}I \right) {\psi }=c_8 {\bar \psi} \left(\Phi'
\right) {\psi} \,, \label{lagrang4} \ee
where $c_8$
is the coupling constant for the octet GB. The GB field $\Phi'$ can be expressed in terms of the GBs and their transition probabilities as \bea
\Phi' = \left( \ba{ccc} P_{\pi}\frac{\pi^0}{\sqrt 2}
+P_{\eta}\frac{\eta}{\sqrt 6}+P_{\eta'}\frac{\eta^{'}}{\sqrt 3} & P_{\pi}\pi^+
& P_K K^+   \\
P_{\pi}\pi^- & -P_{\pi}\frac{\pi^0}{\sqrt 2} +P_{\eta} \frac{\eta}{\sqrt 6}
+P_{\eta'}\frac{\eta^{'}}{\sqrt 3}  &  P_K K^o  \\
P_K K^-  & P_K \bar{K}^0  &  -P_{\eta} \frac{2\eta}{\sqrt 6}
+P_{\eta'}\frac{\eta^{'}}{\sqrt 3} \ea \right). \label{GBfield} \eea
The fluctuation process describing the effective Lagrangian is
\be q^{\uparrow \downarrow} \rightarrow {\rm GB} + q^{'
	\downarrow \uparrow} \rightarrow (q \bar q^{'}) +q^{'	\downarrow \uparrow}\,, \label{basic}\ee
where $q \bar q^{'} +q^{'}$
constitute the sea quarks. In Eq. (\ref{GBfield}), the chiral fluctuations  $u(d) \rightarrow d(u) + \pi^{+(-)}$, $u(d) \rightarrow s + K^{+(0)}$, $u(d,s)\rightarrow u(d,s) + \eta$, and $u(d,s) \rightarrow u(d,s) + \eta^{'}$  are given in terms of the transition probabilities $P_{\pi}$,  $P_K$, $P_{\eta}$ and $P_{\eta'}$ respectively \cite{cheng,johan,hd}. 

From Eq. (\ref{sea-q}) the transition probability of the emission of a GB from
any of the $q$ quark are expressed in terms of the transition probabilities $P_{\pi}$,  $P_K$, $P_{\eta}$ and $P_{\eta'}$ as
\be  P(u,GB) =   P(d,GB)= -\frac{1}{6}\left( 9P_{\pi}+P_{\eta}+2 P_{\eta'}+ 6P_K\right)\,, \ee
\be  P(s,GB)= -\frac{1}{3}\left( 2 P_{\eta}+P_{\eta'}+6 P_K\right)\,, \ee
whereas the transition
probability of the $q^{\uparrow \downarrow}$ quark can be expressed as
\be P(u^{\uparrow \downarrow},GB)=\frac{1}{6}(3 P_{\pi}+P_{\eta}+2P_{\eta'})u^{ \downarrow \uparrow}+
P_{\pi} d^{\downarrow \uparrow}+ P_K s^{\downarrow \uparrow}\,,      \ee

\be P(d^{\uparrow \downarrow},GB)=P_{\pi} u^{\downarrow \uparrow}+
\frac{1}{6}(3P_{\pi}+P_{\eta}+2 P_{\eta'})d^{\downarrow \uparrow}+  P_K s^{\downarrow \uparrow}\,,
\ee

\be P(s^{\uparrow \downarrow},GB)=   P_K u^{\downarrow \uparrow}+ P_K d^{\downarrow \uparrow}+\frac{1}{3}(2 P_{\eta}+P_{\eta'})s^{\downarrow \uparrow}\,.\ee
Using the sea spin polarizations, the magnetic moment contributions coming from the sea quarks can  be calculated from Eq. (\ref{mag-sea}) and have been summarized in Table \ref{sea-trans} for all the decuplet to octet transitions.

\begin{table}
	\tabcolsep -0.1mm {\renewcommand{\arraystretch}{1.5}
		\begin{center}
			\begin{tabular}{|c|c|} \hline 
				$\frac{3}{2}^+ \rightarrow \frac{1}{2}^+$ transition &  Sea quark transition magnetic moments \\ \hline 
							$\mu (\Delta \rightarrow p )_{{\rm S}}$  & $- \frac{2\sqrt{2}}{3}\left(P_{\pi}+ P_K+
				\frac{P_{\eta}}{3}+ \frac{2}{3}P_{\eta'}
				\right)\mu_{u}+ \frac{2\sqrt{2}}{3}\left(P_{\pi}+ P_K+
				\frac{P_{\eta}}{3}+ \frac{2}{3}P_{\eta'}\right)\mu_{d}$ \\\hline
									$\mu (\Sigma^{*+} \rightarrow \Sigma^{+} )_{{\rm S}}$ & $-\frac{2\sqrt{2}}{3}\left( 2 P_{\pi}+
				\frac{P_{\eta}}{3}+ \frac{2}{3}P_{\eta'}
				\right)\mu_{u} -\frac{2\sqrt{2}}{3}\left(P_{\pi}- P_K\right)\mu_{d}
				$ \\ &$ + \frac{2\sqrt{2}}{3} \left(P_K+ \frac{4}{3}P_{\eta}+
			\frac{2}{3}P_{\eta'}\right)\mu_{s}$\\ \hline
				
				$\mu ({\Sigma^{*0}\rightarrow\Sigma^{0}})_{{\rm S}}$ & $-\frac{\sqrt{2}}{3}\left(3 P_{\pi}-
			P_K+ \frac{P_{\eta}}{3}+ \frac{2}{3}P_{\eta'}\right) \mu_{u} -\frac{\sqrt{2}}{3}\left(3P_{\pi}-
				P_K+ \frac{P_{\eta}}{3}+ \frac{2}{3}P_{\eta'}\right)\mu_{d} $ \\ &$ + \frac{2\sqrt{2}}{3}\left(P_K+
				\frac{4}{3}P_{\eta}+ \frac{2}{3}P_{\eta'}\right)\mu_{s}$\\ \hline
				
					$\mu({\Sigma^{*0} \rightarrow \Lambda})_{{\rm S}}$ & $-\sqrt{\frac{2}{3}} \left(P_{\pi}+
				P_K+ \frac{P_{\eta}}{3}+ \frac{2}{3}P_{\eta'}\right)\mu_{u}+ \sqrt{\frac{2}{3}}\left(P_{\pi}+
				P_K+ \frac{P_{\eta}}{3}+ \frac{2}{3}P_{\eta'}\right)\mu_{d}$\\ \hline
				
				$\mu({\Sigma^{*-}\rightarrow\Sigma^{-}})_{{\rm S}}$ & $-\frac{2\sqrt{2}}{3}\left(P_{\pi}-
				P_K\right)\mu_{u}-\frac{2\sqrt{2}}{3}\left(P_{\pi}+
				\frac{P_{\eta}}{3}+ \frac{2}{3}P_{\eta'}\right)\mu_{d} $ \\ &$+ \frac{2\sqrt{2}}{3}a
				\left(P_K+ \frac{4}{3}P_{\eta}+ \frac{2}{3}P_{\eta'}\right)\mu_{s}$\\ \hline
				
				$\mu({\Xi^{*0}\rightarrow\Xi^{0}})_{{\rm S}}$ & $-\frac{2\sqrt{2}}{3}\left(2 P_{\pi}+
				\frac{P_{\eta}}{3}+ \frac{2}{3}P_{\eta'}\right)\mu_{u}- \frac{2\sqrt{2}}{3}\left(P_{\pi}-
				P_K \right)\mu_{d} $ \\ &$ + \frac{2\sqrt{2}}{3}\left(P_K+
				\frac{4}{3}P_{\eta}+ \frac{2}{3}P_{\eta'}\right)\mu_{s}$\\ \hline
				
				$\mu({\Xi^{*-}\rightarrow\Xi^{-}})_{{\rm S}}$ & $- \frac{2\sqrt{2}}{3} \left(2 P_{\pi} + \frac{P_{\eta}}{3}+
				\frac{2}{3}P_{\eta'}\right)\mu_{u} - \frac{2\sqrt{2}}{3}\left(P_{\pi}- P_K
				\right) \mu_{d} $ \\ &$+
				\frac{2\sqrt{2}}{3}\left(P_K+ \frac{4}{3}P_{\eta}+
				\frac{2}{3}P_{\eta'}\right)\mu_{s}$\\ \hline

			\end{tabular}
			
			\caption{Sea quark transition magnetic moments for ${\frac{3}{2}}^+ \to {\frac{1}{2}}^+$ transitions.}
			\label{sea-trans}
	\end{center} }
\end{table}

The magnetic moment contribution of the
angular momentum of a given sea quark can be expressed in terms of 
the orbital angular momenta of quarks and GB $\langle L_q, L_{{\rm GB}} \rangle$ which are further related to the masses of quarks and GB ($M_q$, ${M}_{{\rm GB}}$) as 
\be
\langle L_q \rangle=\frac{{M}_{{\rm GB}}}{M_q+{M}_{{\rm GB}}} ~{\rm and}
~\langle L_{{\rm GB}} \rangle=\frac{M_q}{M_q+{M}_{{\rm GB}}}\,.
\ee
The magnetic moment arising from all the possible transitions of a given valence quark to the GBs is obtained by multiplying the orbital moment of each process to the probability
for such a process to take place. The general orbital moment for any quark $q$ is given as 
\be
\mu (q^{\uparrow} \rightarrow {q}^{'\downarrow}) =\frac{e_{q^{'}}}{2M_q}
\langle L_q \rangle +
\frac{{e}_{q}-{e}_{q^{'}}}{2 {M}_{{\rm GB}}}\langle {L}_{{\rm GB}} \rangle\,. 
\ee
The  orbital moments of $u$, $d$ and $s$ quarks after including the transition probabilities $P_{\pi}$, $P_K$, $P_{\eta}$ and $P_{\eta'}$  as well as the masses of GBs $M_{\pi}$, $M_{K}$
and $M_{\eta}$ can be expressed as 
\be [\mu(u^\uparrow \rightarrow)] =  \left [\frac{3 P_{\pi} M^2_{u}}{2
{M}_{\pi}(M_u+ {M}_{\pi})}- \frac{P_K(M^2_{K}- 3 M^2_{u})}{2
{M}_{K}(M_u+ {M}_{K})} + \frac{P_{\eta} M_{\eta}}{6(M_u+
{M}_{\eta})}+ \frac{P_{\eta'} M_{\eta'}}{3(M_u+ {M}_{\eta'})}  \right]
{\mu}_u \,, \label{orbitu} \ee
\be [\mu(d^\uparrow \rightarrow)] = -
\left [\frac{3 P_{\pi}( M^2_{\pi}-2 M^2_{d})}{2 {M}_{\pi}(M_d+ {M}_{\pi})}-
\frac{P_K M_{K}}{(M_d+ {M}_{K})}  - \frac{P_{\eta}
M_{\eta}}{6(M_d+ {M}_{\eta})}- \frac{P_{\eta'} M_{\eta'}}{3(M_d+
{M}_{\eta'})} \right ] {\mu}_d \,, \label{orbitd} \ee
\be [\mu(s^\uparrow
\rightarrow)] = -  \left[ \frac{P_K (M^2_{K}-3
M^2_s)}{{M}_{K}(M_s+ {M}_{K})}  - \frac{2P_{\eta} M_{\eta}}{3(M_s+
{M}_{\eta})}  - \frac{P_{\eta'} M_{\eta'}}{3(M_s+ {M}_{\eta'})} \right
]{\mu}_s\,. \label{orbits} \ee

The orbital contribution to the magnetic moment of the decuplet to octet transition $\mu\left (B_{\frac{3}{2}^+} \rightarrow B_{\frac{1}{2}^+}\right )_{{\rm O}}$  for the baryon the type
$B(Q_1 Q_2 Q_3)$ is given as
\be
\Delta Q_1\left ({\frac{3}{2}^+} \rightarrow {\frac{1}{2}^+}\right )_{{\rm V}}~\mu(Q_1^{\uparrow} \rightarrow )+\Delta Q_2\left ({\frac{3}{2}^+} \rightarrow {\frac{1}{2}^+}\right )_{{\rm V}}~\mu(Q_2^{\uparrow} \rightarrow )+\Delta Q_3\left ({\frac{3}{2}^+} \rightarrow {\frac{1}{2}^+}\right )_{{\rm V}}~\mu(Q_3^{\uparrow} \rightarrow ) \label{orbital-29}\ee

\section{Results and Discussion}
\label{results}

The transition probabilities $P_{\pi}$, $P_K$, $P_{\eta}$ and $P_{\eta'}$  as well as the masses of GBs $M_{\pi}$, $M_{K}$
and $M_{\eta}$ are the input parameters needed for the numeric
calculations of the baryon transition magnetic moments $\mu\left (B_{\frac{3}{2}^+} \rightarrow B_{\frac{1}{2}^+}\right )$ in the $\chi$CQM. The hierarchy followed by the transition probabilities $P_{\pi}$, $P_K$, $P_{\eta}$ and $P_{\eta'}$ which represent respectively the probabilities of fluctuations of a constituent
quark into pions, $K$, $\eta$ and    $\eta^{'}$ is given as
\be
P_{\eta'}< P_{\eta} < P_K < P_{\pi}\,.
 \ee 
This order is because of the fact that probability of emission of a particular GB is dependent on its masses implying that the probability of emission a heavier meson from a
lighter quark is much smaller than that of the lighter
mesons. The transition probabilities are usually fixed by the experimentally known spin and flavor distribution functions measured from
the DIS experiments \cite{pdg,adams,e866,hd}. A detailed analysis leads to the following probabilities: 
\be P_{\eta'} = 0.03\,, ~~P_{\eta}= 0.04\,,~~P_{K}= 0.06\,,~~  P_{\pi}=0.12\,. \ee
On the other hand, the orbital angular momentum
contributions are characterized by the masses of quarks and GBs ($M_q$ and  $M_{GB}$). The on mass shell mass values can be used in accordance with several other similar calculations \cite{cheng,hdmagnetic}.

\begin{table} 
\hspace {2cm} 	
\tabcolsep 0.5mm 	
{\renewcommand{\arraystretch}{1.4}
\hbox{			
\begin{sideways}	
\begin{tabular}{|c|c|c|c|c|c|}      \hline 
					
				$B_{\frac{3}{2}^+} \rightarrow B_{\frac{1}{2}^+}$  & Data \cite{pdg}  & 	$\mu\left (B_{\frac{3}{2}^+} \rightarrow B_{\frac{1}{2}^+}\right )_{{\rm V}}$ &$\mu\left (B_{\frac{3}{2}^+} \rightarrow B_{\frac{1}{2}^+}\right )_{{\rm S}}$ &$\mu\left (B_{\frac{3}{2}^+} \rightarrow B_{\frac{1}{2}^+}\right )_{{\rm O}}$ & $\mu\left (B_{\frac{3}{2}^+} \rightarrow B_{\frac{1}{2}^+}\right )$ \\ 
					 Transition & & 	 & & &  \\ \hline 
				
				$\Delta \rightarrow p $  & $-3.43$  & 2.83&$-0.60$& 1.65& 3.87 \\
				
				$\Sigma^{*+} \rightarrow \Sigma^{+} $   & 4.45& 2.49&$-0.54$& 0.65& 2.60 \\
				
				${\Sigma^{*0}\rightarrow\Sigma^{0}}$   &$--$ &1.08& $-0.24$ & $0.18$& 1.02 \\
				
					${\Sigma^{*0} \rightarrow \Lambda}$   & 3.69& 2.45&$-0.52$ & 1.42& 3.35 \\ 
				${\Sigma^{*-}\rightarrow\Sigma^{-}}$  &$< 0.85$ & $-0.33$& $0.05$ & $-0.99$& $-1.27$ \\
				
				${\Xi^{*0}\rightarrow\Xi^{0}}$   &$<5.39$ & 2.50& $-0.31$& $0.65$ &  $2.83$  \\
				
				${\Xi^{*-}\rightarrow\Xi^{-}}$   &$<5.39$ &$-0.33$& $0.31$ & $-1.00$& $-1.02$ \\

				\hline
				\end{tabular}
			\end{sideways}
			\hspace {0.6cm}
			\rotcaption{Magnetic moments in units of $\mu_N$ for the $B_{\frac{3}{2}^+} \rightarrow B_{\frac{1}{2}^+}$ transitions.} \label{transitions}}}
\end{table}

The inputs discussed  above are used to calculate the explicit valence $\mu\left (B_{\frac{3}{2}^+} \rightarrow B_{\frac{1}{2}^+}\right )_{{\rm V}}$, sea $\mu\left (B_{\frac{3}{2}^+} \rightarrow B_{\frac{1}{2}^+}\right )_{{\rm S}}$ and orbital $\mu\left (B_{\frac{3}{2}^+} \rightarrow B_{\frac{1}{2}^+}\right )_{{\rm O}}$ contributions corresponding to the transition magnetic moments and the results have been presented in Table  \ref{transitions}. The limited experimental data available for the ${\frac{3}{2}^+} \rightarrow {\frac{1}{2}^+}$ transitions has also been presented in the table. It can immediately seen that the contributions coming from valence and orbital contributions  have same signs whereas the sea contributions have opposite signs. All of ultimately add up to give the total magnetic moment.   It is also observed that in some cases the orbital part dominates over the sea part making the total magnetic moments even higher than the valence part. This is for the case of  	$\Delta \rightarrow p $, $\Sigma^{*+} \rightarrow \Sigma^{+} $ , ${\Sigma^{*0} \rightarrow \Lambda}$,  ${\Sigma^{*-}\rightarrow\Sigma^{-}}$ 	${\Xi^{*-}\rightarrow\Xi^{-}}$ and ${\Xi^{*0}\rightarrow\Xi^{0}}$  transitions. 
One can generalize this as follows, whenever there the number of $u$, $d$ or $s$ quarks are more, there is dominance of the orbital part, whereas when the $u$, $d$ and $s$ quarks are in equal numbers,
there is some variation from this behavior. For example, in the case of $\Delta \rightarrow p $  and
$\Sigma^{*+} \rightarrow \Sigma^{+} $  the baryon quark content is $uud$ and clearly the orbital contribution is dominant as compared to the sea contribution. Similarly, in the cases of 	${\Sigma^{*-}\rightarrow\Sigma^{-}}$, ${\Xi^{*0}\rightarrow\Xi^{0}}$   and ${\Xi^{*-}\rightarrow\Xi^{-}}$ transitions, where the baryon quark contents are $dds$, $uss$ and $dss$ respectively, again the orbital contributions are large as compared to the sea contributions.
This can be easily explained from Eq. (\ref{orbital-29}) where we can see that the orbital contribution is governed by the valence quarks spin polarization alongwith the orbital moments of the quarks. The orbital part dominates in the cases where the number of $u$, $d$ or $s$ quarks are more 
because the contributions of the individual quarks with the same magnitudes add up in the same direction. On the other hand, in the cases where the quark content is $uds$ (${\Sigma^{*0}\rightarrow\Sigma^{0}}$   and ${\Sigma^{*0} \rightarrow \Lambda}$ transitions)  there is some variation from this behavior for different cases because the contributions of the individual quarks have different magnitudes.
These observations clearly suggest that since the quark sea is created from the quantum fluctuation
associated with the bound state hadron dynamics and the
process is completely determined by nonperturbative
mechanisms, the constituent quarks and weakly interacting Goldstone bosons can
provide the appropriate degree of freedom in the
nonperturbative regime of QCD on which further corrections could be evaluated. A further precise measurement of these magnetic moments, therefore, would have great importance for the understanding of $\chi$CQM.

For the sake of comparison with other models, we have presented the results of the available phenomenological and theoretical models in Table \ref{data}. We have presented the results from chiral perturbation theory ($\chi$PT) \cite{chipt-transition}, light cone QCD sum rules  and light cone QCD (LCQCDSR) \cite{aliev-transition}, large-$N_c$ chiral perturbation theory (Large $N_C$ PT) \cite{1largeNc-transition,2largeNc-transition}, relativistic quark model (Rel-QM)\cite{rel-transition},     QCD sum rules (QCDSR) \cite{1qcdsr-transition,2qcdsr-transition},  lattice
QCD \cite{lattice-transition}, chiral quark model ($\chi$QM) \cite{chiqm-transition} effective mass quark model (EMQM) \cite{emqm-transition}, meson cloud model (MCM) \cite{meson-cloud-transition} , U-spin \cite{uspin-transition}.  
The experimental values from the PDG \cite{pdg} have also been listed.

\begin{table}
\tabcolsep 3.5mm
\hbox{
\begin{sideways}{\footnotesize
\begin{tabular}{|l|c|c|c|c|c|c|c|} \hline  Other models  & 	$\Delta \rightarrow p $  &  
	
	$\Sigma^{*+} \rightarrow \Sigma^{+} $   &
	${\Sigma^{*0}\rightarrow\Sigma^{0}}$   &
		${\Sigma^{*0} \rightarrow \Lambda}$ &

	${\Sigma^{*-}\rightarrow\Sigma^{-}}$  & 
	
	${\Xi^{*0}\rightarrow\Xi^{0}}$   & 
	
	${\Xi^{*-}\rightarrow\Xi^{-}}$   
	
  \\  \hline 

NQM \cite{nqm} & 2.65 & 2.42 & 1.05 & 2.31 & $-0.32$ & 2.18 &$-0.29$ \\

$\chi$PT \cite{chipt-transition} & $-3.50$ & 4.46 &$-2.34$ & 3.62& $-0.21$ & 5.38& 0.20 \\

LCQCDSR \cite{aliev-transition}& $2.50$ & 2.10 &$0.89$ & $--$& $-0.47$ & $-2.77$& $0.47$ \\

Large $N_C$ PT \cite{1largeNc-transition} & $3.51$ & 2.96 &$1.34$ & 2.96& $-0.27$ & 2.96& $--$ \\

Large $N_C$ PT \cite{2largeNc-transition}& $3.51$ & 2.97 &$1.39$ & 2.93& $-0.19$ & 2.96& $-0.19$ \\

Rel-QM \cite{rel-transition}& $3.25$ & 2.59 &$1.07$ & 2.86& $-0.46$ & 2.71& $-0.47$ \\

QCDSR \cite{1qcdsr-transition} &$-2.76$&2.24&1.01&$-2.46$& $-0.22$ &$2.46$ & $-0.27$ \\

QCDSR \cite{2qcdsr-transition}&$3.86$&3.38&1.47&$4.44$& $-0.57$ &$-1.24$ & $0.23$ \\

Lattice QCD \cite{lattice-transition} &2.46&2.61&1.07&$--$& $-0.47$ &$-2.77$ & 0.47 \\

$\chi$QM \cite{chiqm-transition} & $-3.31$ & 2.17 &$--$ & $-2.74$& $-0.59$ & 2.23& $-0.59$ \\

EMQM \cite{emqm-transition} & $2.63$ & 2.33 &$1.02$ & 2.28& $0.30$ & 2.33& $0.30$ \\

MCM \cite{meson-cloud-transition} & $3.32$ & 3.54 &$1.61$ & 3.39& $-0.34$ & 3.62& $-0.42$ \\

U-spin \cite{uspin-transition}  & $--$ & 3.22 &$1.61$ & 2.68& $0$ & 3.21& $--$ \\

PDG \cite{pdg}    & $-3.43$ & 4.45 &$--$ & 3.69& $< 0.85$ & $<5.39$& $<5.39$ \\
 \hline
This work: $\chi$CQM & 3.87& 2.60& 1.02& 3.35&$-1.27$& 2.83& $-1.02$ \\ \hline  \end{tabular}}

\end{sideways}
\hspace {0.6cm}
\rotcaption{Phenomenological results of some other theoretical
approaches for $B_{\frac{3}{2}^+} \rightarrow B_{\frac{1}{2}^+}$ transition magnetic moments.} \label{data}}
\end{table}

Since the experimental data is available for $\Delta \rightarrow p $, $\Sigma^{*+} \rightarrow \Sigma^{+} $, and ${\Sigma^{*0} \rightarrow \Lambda}$ transitions, we can compare these with the $\chi$CQM results as well as with the results of other available models. It is evident from
the table that a good agreement corresponding to the case of  $\Delta \rightarrow p $ is obtained. 
The magnetic moment of $\Delta \to p + \gamma$ transition is a long standing
problem and most of the approaches in literature underestimate it. The empirical estimate for the magnetic moment of the $\Delta \to p +
\gamma$ transition can be made from the helicity amplitudes \cite{pdg},
$A_{\frac{1}{2}}$ = $-$ 0.135 $\pm$ 0.005 GeV$^{-\frac{1}{2}}$ and
$A_{\frac{3}{2}}$ = $-$ 0.250 $\pm$ 0.008 GeV$^{-\frac{1}{2}}$
\cite{pdg} as inputs in the decay rate. The extracted magnetic moment comes out to be
$\mu_{\Delta \rightarrow p}$ = 3.46 $\pm$ 0.03 $\mu_N$. Our predicted value of 3.87 $\mu_N$ is very close to the experimental results. The difference in sign with some of the other models may be due to the different model predictions and signs of the wavefunctions.  In the case of $\Sigma^{*+} \rightarrow \Sigma^{+} $, even though our results are almost half of the experimental value,  except for a very few models all other models predict a value close to our results.  
In the case of and ${\Sigma^{*0} \rightarrow \Lambda}$ transitions  our results are more or less in good agreement with the results of other models as well as with the experimental data.

To summarize, the  chiral constituent quark model ($\chi$CQM) is able to describe the transition properties of the low lying baryons. In a very interesting manner, the $\chi$CQM) is able to phenomenologically estimate the explicit contributions coming from the valence quarks, sea quarks and their orbital angular momentum to the total magnetic moments of the transitions from the $J^P=\frac{3}{2}^+$ decuplet to $J^P=\frac{1}{2}^+$ octet baryons. The results immediately suggest that the contributions coming from various sources with  same and opposite signs ultimately add up to give the total magnetic moment. 
Whenever the number of $u$, $d$ or $s$ quarks are more, there is a dominance of the orbital part, whereas when the $u$, $d$ and $s$ quarks are in equal numbers,
there is some variation from this behavior. These observations endorse that the
sea quarks and the orbital angular momentum of the sea quarks perhaps provide the dominant dynamics of the
constituents in the low-energy regime of QCD. The qualitative and quantitative description of the results confirms that constituent quarks and weakly interacting Goldstone bosons provide the appropriate degree of freedom in the nonperturbative regime of QCD. A further precise measurement of these magnetic moments, therefore, would have
important implications for the $\chi$CQM.


\end{document}